\documentclass[12pt]{article}
\usepackage{amsmath,amssymb,latexsym,graphicx}

\begin{document}
\begin{flushright}\small
 \begin{minipage}[b]{3cm}
  {YITP-06-35\\KUNS-2032\\OIQP-06-09\\hep-th/0607182}
 \end{minipage}
\end{flushright}

\begin{center}
 {\Large\bf Dirac Sea and Hole Theory for Bosons II}\\[1em]
 {\large\it --- Renormalization Approach ---}\\[2em]
{%
Yoshinobu \textsc{Habara}$^{1}$, %
Yukinori \textsc{Nagatani}$^{2}$,\\%
Holger B. \textsc{Nielsen}$^{3}$ %
and Masao \textsc{Ninomiya}$^{4}$\footnotemark}\\[1em]
{\it%
$^1$Department of Physics, Graduate School of Science,\\%
Kyoto University, Kyoto 606-8502, Japan\\[0.5em]%
$^2$Okayama Institute for Quantum Physics,\\%
Kyoyama-cho 1-9, Okayama-city 700-0015, Japan\\[0.5em]%
$^3$Niels Bohr Institute, University of Copenhagen,\\%
17 Blegdamsvej Copenhagen \o, Denmark\\[0.5em]%
$^4$Yukawa Institute for Theoretical Physics,\\%
Kyoto University, Kyoto 606-8502, Japan%
}
\end{center}

\begin{abstract}
{%
In bosonic formulation of the negative energy sea,
so called Dirac sea presented in the preceding paper
[arXiv:hep-th/0603242],
one of the crucial points is how to construct a positive definite inner
product in the negative energy states,
since naive attempts would lead to non-positive definite ones.
In the preceding paper the non-local method is used
to define the positive definite inner product.
In the present article
we make use of a kind of $\epsilon$-regularization
and renormalization method which may clarify transparently
the analytical properties of our formulation.
}
\end{abstract}

\footnotetext{Also working at Okayama Institute for Quantum Physics, Kyoyama-cho 1-9, Okayama-city 700-0015, Japan}

\section{Introduction}

Recently, a long-standing problem or puzzle \cite{Weinberg:1995mt}
in any relativistic quantum field theories has been investigated by the 
present authors
\cite{Nielsen:1998mc,Habara:2003cz,nn3,Habara:2006eg}.
The problem is how to construct the negative energy sea, or Dirac sea
for bosons, since as is well known the fermion fields was
historically second quantized firstly by Dirac in terms of Dirac sea
and hole theory \cite{dirac}.
In the fermion case there exists Pauli's exclusion principle
and easily negative energy sea, namely, the Dirac sea is constructed.
In the bosonic cases contrary to fermions,
one might think at first that
it would be impossible to construct such a sea
due to lack of the Pauli principle,
so that infinite number of bosons at each energy state could exist
and thus the negative energy states could never be filled.
However,
we succeeded in constructing the Dirac sea for bosons,
so called {\em boson sea}.
In fact, there we solved one of the serious problems:
how to construct the positive definite norm
of the negative energy states.
There we have used a non-local definition
(the detail of the methods see \cite{Habara:2006eg}).

It is the purpose of the present article to show another method
employing the regularization of the naively divergent inner product
in the negative number sector and the renormalization.
In fact we make use of a kind of $\epsilon$-regularization method to make
it finite,
and then make renormalization by discarding all the divergent terms,
which can be done successfully.
The advantage of this $\epsilon$-method is to make transparent
the analytic structure of the whole procedure.

The present paper is organized as follows:
In the following section 2 we treat the inner product by
$\epsilon$-regularization and renormalization,
and the positive definite inner product is obtained.
In section 3 we verify the orthonormality of our inner product
obtained in section 2 by performing some explicit analytic calculation.
In section 4 we present another definition of the inner product
without the subtraction scheme in the renormalization,
and we give a proof of the orthonormality.
Section 5 is devoted to conclusion and further perspectives.

\section{Inner Product by Renormalization}

\begin{figure}[htbp]
 \begin{center}
  \includegraphics{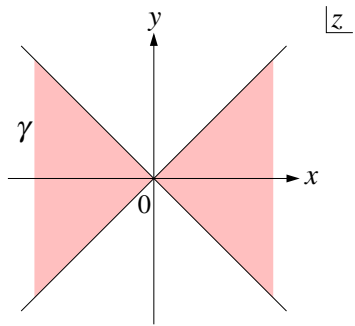}
 \end{center}
 \caption[region]{%
 The region $\gamma$ on $(x,y)$-plane.
 The exponential factor $e^{-\frac{1}{2}(x^2 - y^2)}$
 converges in this region.
 }
 \label{gamma}
\end{figure}

As a preparation to define the inner product, 
we define an $\varepsilon$-regularized inner product as
\begin{eqnarray}
 \langle f \:|\: g \rangle_\varepsilon
  &=&
 \int_{\gamma} dx \: dy \>
  \left<
   f (x,y),\:
   g (x,y)
  \right> \Lambda_\epsilon(x),
 \label{pre-gamma-product}
\end{eqnarray}
where
\begin{eqnarray}
 \gamma
  &\equiv&
  \left\{ \: (x,\:y) \;\Big|\;
   x^2 - y^2 \geq 0
  \right\}
  \label{gamma-region}
\end{eqnarray}
is the integral region, and
\begin{eqnarray}
 \Lambda_\varepsilon(x)
  &\equiv&
  \frac{1}{-\log\varepsilon} e^{-\varepsilon x^2}
  \label{reg-func}
\end{eqnarray}
is a regularization function.
The integral region $\gamma$ is just the inside of the light-cone
shown as the shaded zone in Fig.~\ref{gamma}.
%
%

The $\varepsilon$-regularized inner product (\ref{pre-gamma-product})
is divergent for $\varepsilon\to0$,
and it is divided into the following three parts:
\begin{eqnarray}
 \langle f \:|\: g \rangle_\varepsilon
  &=&
  (\varepsilon\mbox{-divergent part}) \;+\;
  (\varepsilon\mbox{-independent part}) \;+\;
  (\varepsilon\mbox{-zero part})
  \label{separation}
\end{eqnarray}
by the behavior of $\varepsilon\to0$.
The first term $(\varepsilon\mbox{-divergent part})$ diverges
for $\varepsilon\to0$.
According to the precise calculation presented in the next section,
the concrete form of $(\varepsilon\mbox{-divergent part})$ is
given by the linear combination of
\begin{eqnarray}
 \frac{1}{-\log\varepsilon} \left(\frac{1}{\varepsilon}\right)^n
\end{eqnarray}
for positive integer n.
Thus we can manifestly separate the second term
$(\varepsilon\mbox{-independent part})$
which is just independent term of $\varepsilon$
for $\varepsilon\to0$.
The third term $(\varepsilon\mbox{-zero part})$ goes to zero
for $\varepsilon\to0$.

We define the inner product by a renormalization of the
$\varepsilon$-regularized inner product:
\begin{eqnarray}
 \langle f \:|\: g \rangle
  &\equiv&
  (\varepsilon\mbox{-independent part})\mbox{ of }
  \langle f \:|\: g \rangle_\varepsilon,
 \label{gamma-product}
\end{eqnarray}
which may be so called the minimal subtraction scheme
of the renormalization.
We can confirm
the inner product (\ref{gamma-product})
satisfies the orthonormal condition:
\begin{eqnarray}
 \left< n_+, -m_- | n_+', -m_-' \right>
  &=&  \delta_{n,n'} \delta_{m,m'}.
  \label{ortho-normal-condition}
\end{eqnarray}
The product (\ref{gamma-product}) is just positive definite
even for the indefinite metric 
of $I,J$-algebra, namely, $\left<J,J\right>=-1$.
Therefore
we construct the Hilbert space including the negative number sector
by using the product (\ref{gamma-product}).
These are the result from 
the combination of the $\gamma$-restriction,
the regularization function $\Lambda_\varepsilon(x)$
and the renormalization.

In the definition of the inner product (\ref{gamma-product}),
the restriction of the integral region into $\gamma$
and the regularization function $\Lambda_\varepsilon(x)$
are quite important.
The restriction of the integral region into $\gamma$
is a part of the regularization
which is nothing but a kind of hard cut-off.
The choice of the integral region $\gamma$ is important
to realize the orthogonal condition.

\section{A Proof of the Orthonormality of the Inner Product}

\begin{figure}[htb]
 \begin{center}
  \includegraphics{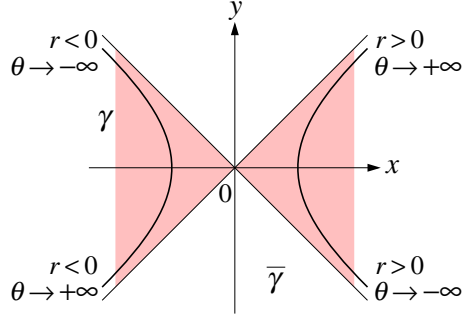}
 \end{center}
 \caption[region2]
 {Hyperbolic coordinate $(r,\theta )$ for fixed $|r|$.}
 \label{hyperbola}
\end{figure}

We hereby verify the orthonormality (\ref{ortho-normal-condition}) of the
inner product (\ref{gamma-product}).
We introduce a hyperbolic coordinate $(r,\theta)$ on the $(x,y)$-plane
for convenience.
The hyperbolic coordinate $(r,\theta)$ on the $(x,y)$-plane
is defined as
\begin{eqnarray}
 &&
  \left\{
   \begin{array}{l}
    x \;=\; r \cosh \theta,\\
    y \;=\; r \sinh \theta ,
   \end{array}
  \right. \label{3.36}
\end{eqnarray}
where $r\in (-\infty ,+\infty)$ and $\theta \in (-\infty ,+\infty)$
covers the whole region $\gamma$ as shown in Fig.~\ref{hyperbola}.
This hyperbolic coordinate respects
the Lorentz invariance of the Hamiltonian 
and the ``Gaussian'' factor $e^{-\frac{1}{2}(x^2 - y^2)}$
(see our previous paper \cite{Habara:2006eg}).
The relation between the differential operators is given by
\begin{eqnarray*}
	\left(
	\begin{array}{c}
	\frac{\partial}{\partial x}\\[0.5em]
	\frac{\partial}{\partial y}
	\end{array}
	\right)
	&=&
	\left(
	\begin{array}{cc}
	+\cosh\theta & -\frac{1}{r}\sinh\theta\\[0.5em]
	-\sinh\theta & +\frac{1}{r}\cosh\theta
	\end{array}
	\right)
	\left(
	\begin{array}{c}	
	\frac{\partial}{\partial r}\\[0.5em]
	\frac{\partial}{\partial\theta}
	\end{array}
	\right),
\end{eqnarray*}
and the relation between integral measures becomes
\begin{eqnarray*}
  \int_{\gamma} dx \: dy
   &=&
   \int_{-\infty}^{+\infty} |r| dr
   \int_{-\infty}^{+\infty} d\theta.
\end{eqnarray*}

As the first step of the proof,
we concretely calculate several inner products and norms of the states.

The most important is the vacuum norm.
The $\varepsilon$-regularized product (\ref{pre-gamma-product})
of the vacuum becomes
\begin{eqnarray}
 \langle 0_+,0_-|0_+,0_- \rangle_{\varepsilon}
  &=&
  \langle I,I \rangle \times
  \frac{1}{-\log\varepsilon}
  \int_{-\infty}^{+\infty} dr
  \int_{-\infty}^{+\infty} d\theta \:
  |r| \: e^{-r^2 \left(1 + \varepsilon\cosh^2\theta\right)}
  \nonumber \\
 &=&
  \frac{1}{-\log\varepsilon}
  \int_{-\infty}^{+\infty}d\theta \> 
  \frac{1}{1+\varepsilon \cosh^2\theta} \nonumber \\
 &=&
  \frac{1}{-\log\varepsilon}
  \frac{\log
  \left(\sqrt{\varepsilon +1}+1\right) 
  -\log\left(\sqrt{\varepsilon +1}-1\right)}{\sqrt{\varepsilon +1}} 
  \nonumber \\
 &=&
  1
  \:-\: \frac{\log 4}{\log\varepsilon}
  \:-\: \frac{1}{2} \varepsilon
  \:-\: \frac{1 - \log 4}{2\log\varepsilon} \varepsilon
  \:-\: \frac{3}{8} \varepsilon^2
  \:+\: \cdots.
  \label{pre-vacuum-norm}
\end{eqnarray}
Then we obtain $\langle 0_+,0_-|0_+,0_- \rangle = 1$,
because
there arises no divergent part in (\ref{pre-vacuum-norm}) and
the $\varepsilon$-independent part of (\ref{pre-vacuum-norm}) is $1$
as $\varepsilon \rightarrow 0$.

We also calculate a product of the orthogonal states
\begin{eqnarray}
 \langle 0_+,0_-|0_+,-1_- \rangle_{\varepsilon}
  &=&
  \langle I,J \rangle \times
  \frac{1}{-\log\varepsilon}
  \int_{-\infty}^{+\infty} dr
  \int_{-\infty}^{+\infty} d\theta \:
  |r|
  \sqrt{2} r \sinh\theta
  \: e^{-r^2 \left(1 + \varepsilon\cosh^2\theta\right)}
  \nonumber\\
 &=& 0.
  \label{pre-01-norm}
\end{eqnarray}
In this case, the orthogonal relation is realized
without any regularization.

One of the non-trivial cases is
\begin{eqnarray}
 \langle 0_+,0_-|0_+,-2_- \rangle_{\varepsilon}
 &=&
  \langle I,J^2 \rangle \times
  \frac{1}{-\log\varepsilon}
 \int_{-\infty}^{+\infty}dr
 \int_{-\infty}^{+\infty}d\theta \:
 |r|
 \left(2r^2\sinh^2\theta + 1\right)
 e^{-r^2(1+\epsilon \cosh^2\theta )}
 \nonumber\\
 &=&
  \frac{-1}{-\log\varepsilon}
  \int_{-\infty}^{+\infty}d\theta \> 
  \left\{
     \frac{2\sinh^2\theta}{(1+\epsilon \cosh^2\theta )^2}
   + \frac{1}{1+\epsilon \cosh^2\theta}
  \right\}
  \nonumber\\
 &=&
  \frac{-2}{-\epsilon\log\epsilon}.
  \label{pre-02-norm}
\end{eqnarray}
Thus
the renormalized product defined in (\ref{gamma-product})
becomes $\langle 0_+,0_-| 0_+,-2_- \rangle = 0$,
because there appears no $\varepsilon$-independent term
rather than the divergent term.

More important case is the following:
\begin{eqnarray}
 \langle 0_+,-1_-|0_+,-1_- \rangle_{\varepsilon}
 &=&
  \langle J,J \rangle \times
  \frac{1}{-\log\varepsilon}
 \int_{-\infty}^{+\infty}dr
 \int_{-\infty}^{+\infty}d\theta \:
 |r|
 \left(2r^2\sinh^2\theta\right)
 e^{-r^2(1+\epsilon \cosh^2\theta )}
 \nonumber\\
 &=&
  \frac{-1}{-\log\varepsilon}
  \int_{-\infty}^{+\infty}d\theta \> 
     \frac{2\sinh^2\theta}{(1+\epsilon \cosh^2\theta )^2}
  \nonumber\\
 &=&
  \frac{-2}{-\epsilon\log\epsilon}
  \:+\:
  \langle 0_+,0_-|0_+,0_- \rangle_{\varepsilon},
  \label{pre-11-norm}
\end{eqnarray}
where we have used the calculation steps (\ref{pre-vacuum-norm})
and (\ref{pre-02-norm}).
We obtain $\langle 0_+,-1_-|0_+,-1_- \rangle = 1$.
We have the positive value of this product
even for the negative norm of the $J$-element.

As the second step of the proof,
we derive recurrence formulae
for the $\varepsilon$-regularized inner product.
For any wave-functions $f(x,y)$ and $g(x,y)$,
we have the following relations:
\newcommand{\ST}{{\cal S}}
\begin{eqnarray}
  \bigl\langle f \:\bigl|\: {\textstyle\frac{d}{dx}g} \bigr\rangle_\varepsilon
  &=&
  \;-\;
  \bigl\langle {\textstyle\frac{d}{dx}f} \:\bigl|\: {g}
  \bigr\rangle_\varepsilon
  \;+\; 
  \ST_x\left[ \left<f,g\right> \right]_\varepsilon
  \;+\;
  2 \varepsilon
  \langle {f} \:| x |\: {g} \rangle_\varepsilon,
  \label{dif-x}
  \\
  \bigl\langle f \:\bigl|\: {\textstyle\frac{d}{dy}g} \bigr\rangle_\varepsilon
  &=&
  \;-\;
  \bigl\langle {\textstyle\frac{d}{dy}f} \:\bigl|\: {g}
  \bigr\rangle_\varepsilon
  \;+\; 
  \ST_y\left[ \left<f,g\right> \right]_\varepsilon,
  \label{dif-y}
\end{eqnarray}
where we have defined surface terms as
\begin{eqnarray}
 \ST_x\left[ \left<f,g\right> \right]_\varepsilon
  &=&
  +
  \frac{1}{-\log\varepsilon}
  \left[
   \int_{-\infty}^{+\infty} d\theta \: |r| \cosh\theta
   \cdot \left<f,g\right>
   \cdot e^{-\epsilon r^2 \cosh^2 \theta}
  \right]_{r=-\infty}^{r=+\infty}
  \nonumber\\
 && 
  -
  \frac{1}{-\log\varepsilon}
  \left[
   \int_{-\infty}^{+\infty} dr \: |r| \frac{\sinh\theta}{r}
   \cdot \left<f,g\right>
   \cdot e^{-\epsilon r^2 \cosh^2 \theta}
  \right]_{\theta =-\infty}^{\theta =+\infty},
  \\
 \ST_y\left[ \left<f,g\right> \right]_\varepsilon
  &=&
  -
  \frac{1}{-\log\varepsilon}
  \left[
   \int_{-\infty}^{+\infty} d\theta \: |r| \sinh\theta
   \cdot \left<f,g\right>
   \cdot e^{-\epsilon r^2 \cosh^2 \theta}
  \right]_{r=-\infty}^{r=+\infty}
  \nonumber\\
 && 
  +
  \frac{1}{-\log\varepsilon}
  \left[
   \int_{-\infty}^{+\infty} dr \: |r| \frac{\cosh\theta}{r}
   \cdot \left<f,g\right>
   \cdot e^{-\epsilon r^2 \cosh^2 \theta}
  \right]_{\theta =-\infty}^{\theta =+\infty}.
\end{eqnarray}
We postpone presenting details of the derivation (\ref{dif-x}) to APPENDIX.
The third term in (\ref{dif-x}) comes from the regularization function
$\Lambda_\varepsilon(x)$ in the definition of the
$\varepsilon$-regularized product (\ref{pre-gamma-product}).
By applying the relation (\ref{dif-x})
into the creation operator
$a_+^\dagger = I\otimes\frac{1}{\sqrt{2}}
\left( x - \frac{\partial}{\partial x}\right)$
of the positive number sector,
we obtain the following relation:
\begin{eqnarray}
  \langle
  \phi_{n_+,-m_-} \:|\: a_+^\dagger \phi_{n'_+ - 1,-m'_-}
  \rangle_\varepsilon
 &=&
  \langle
  a_+ \phi_{n_+,-m_-} \:|\: \phi_{n'_+ - 1,-m'_-}
  \rangle_\varepsilon
  \nonumber\\
 &&
  \;-\;
  \frac{1}{\sqrt{2}}
  \ST_x
  \left[ \langle\phi_{n_+,-m_-},\:\phi_{n'_+ - 1,-m'_-}\rangle
  \right]_\varepsilon
  \nonumber\\
 &&
  \;-\;
  \sqrt{2} \varepsilon
  \langle \phi_{n_+,-m_-} \:|\:x\: \phi_{n'_+-1,-m'_-} \rangle_\varepsilon.
  \label{aa-x}
\end{eqnarray}
By using the relation $I\otimes\sqrt{2} x = a_+ + a_+^\dagger$
in the third term in (\ref{aa-x})
and by operating the creation and annihilation operators
on the wave functions,
the relation (\ref{aa-x}) becomes a relation among three energy levels:
\begin{eqnarray}
 \sqrt{n'_+} (1+\varepsilon)
 \langle \phi_{n_+,-m_-} \:|\: \phi_{n'_+,-m'_-} \rangle_\varepsilon
  &=&
  \sqrt{n_+}
  \langle
  \phi_{n_+ - 1,-m_-} \:|\: \phi_{n'_+ - 1,-m'_-}
  \rangle_\varepsilon
  \nonumber\\
 &&
  \;-\;
  \frac{1}{\sqrt{2}}
  \ST_x
  \left[ \langle\phi_{n_+,-m_-},\:\phi_{n'_+ - 1,-m'_-}\rangle
  \right]_\varepsilon
  \nonumber\\
 &&
  \;-\;
  \sqrt{n'-1}
  \varepsilon
  \langle
  \phi_{n_+,-m_-} \:|
  \: \phi_{n'_+-2,-m'_-}
  \rangle_\varepsilon.
  \label{level-x}
\end{eqnarray}
We can derive the similar relation to (\ref{aa-x}) and (\ref{level-x})
for the negative number sector.
This derivation is simpler than that of the positive number sector,
due to the absence of the third term in (\ref{dif-y}).
In fact
by using the relation (\ref{dif-y}) in the annihilation operator
$a_- = J\otimes\frac{1}{\sqrt{2}}
\left( y + \frac{\partial}{\partial y}\right)$,
we obtain
\begin{eqnarray}
  \langle
  \phi_{n_+,-m_-} \:|\: a_- \phi_{n'_+,-m'_- + 1}
  \rangle_\varepsilon
 &=&
  \langle
  a_-^\dagger \phi_{n_+,-m_-} \:|\: \phi_{n'_+,-m'_- + 1}
  \rangle_\varepsilon
  \nonumber\\
 &&
  \;+\;
  \frac{1}{\sqrt{2}}
  \ST_y
  \left[ \langle\phi_{n_+,-m_-}, \: J \: \phi_{n'_+,-m'_- + 1}\rangle
  \right]_\varepsilon,
  \label{aa-y}
\end{eqnarray}
where we have used the property
$\left< f,\: J g \right> = \left<J f,\: g \right> $.
By the operation of the creation and annihilation operators,
the relation (\ref{aa-y}) becomes
\begin{eqnarray}
 \sqrt{m'_-}
 \langle \phi_{n_+,-m_-} \:|\: \phi_{n'_+,-m'_-} \rangle_\varepsilon
  &=&
  \sqrt{m_-}
  \langle
  \phi_{n_+,-m_- + 1} \:|\: \phi_{n'_+,-m'_- + 1}
  \rangle_\varepsilon
  \nonumber\\
 &&
  \;+\;
  \frac{1}{\sqrt{2}}
  \ST_y
  \left[ \langle\phi_{n_+,-m_-},\:J\:\phi_{n'_+,-m'_- + 1}\rangle
  \right]_\varepsilon.
  \label{level-y}
\end{eqnarray}
The equations (\ref{level-x}) and (\ref{level-y})
are recurrence formulae which determine the
concrete values of the $\varepsilon$-regularized inner product.
The first terms of the recurrence formulae are given by 
the $\varepsilon$-regularized norm of vacuum in (\ref{pre-vacuum-norm}), 
namely,
\begin{eqnarray}
 \langle 0_+,0_-|0_+,0_- \rangle_{\varepsilon}
  &=&
  1 \;+\; (\varepsilon\mbox{-zero part}),
\end{eqnarray}
and the properties
\begin{eqnarray}
 &&
  a_+ |0_+,-m_- \rangle = 0,
  \qquad
  a_-^\dagger |n_+,0_- \rangle = 0.
\end{eqnarray}

We consider $\varepsilon$-behavior of the surface term
in the recurrence formula (\ref{level-x}). 
The surface term for any $n_+,\:m_-,\:n'_+$ and $m'_-$
is given by linear combination:
\begin{eqnarray}
 \ST_x
  \left[
   \langle\phi_{n_+,-m_-},\:\phi_{n'_+ - 1,-m'_-}\rangle
  \right]_\varepsilon
  &=&
  {
  \sum_{a=0}^{n_++n_+'-1} \;
  \sum_{b=0}^{m_-+m_-'}}
  C_{a,b} \:
  \ST_x
  \left[ x^a y^b e^{-x^2 + y^2} \right]_\varepsilon,
  \label{linear-combination-x}
\end{eqnarray}
where $C_{a,b}$ is a coefficient of the linear combination,
because
$\ST_x[\cdot]_\varepsilon$ is a linear functional and
the product
$\langle\phi_{n_+,-m_-},\:\phi_{n'_+ - 1,-m'_-}\rangle$
is obtained by the linear combination of the functions;
$x^a y^b e^{-x^2 + y^2}$ for integers $a,b \geq 0$.
In the last term of (\ref{linear-combination-x}) reads
\begin{eqnarray}
 \ST_x
  \left[ x^a y^b e^{-x^2 + y^2} \right]_\varepsilon
  &=&
  \frac{-2}{-\log\varepsilon}
  \:
  \left(\frac{1}{\varepsilon}\right)^{\frac{a+b+1}{2}}
  \: \times
  \left\{
   \begin{array}{ll}
     1 & (a=1,\ b=0)\\
     \frac{a+b-1}{2}\Gamma\left(\frac{a+b-1}{2}\right)
       & (a:{\rm odd},\ b:{\rm even})\\
     0 & ({\rm others})
   \end{array}
  \right.,
\end{eqnarray}
which is zero or diverges for $\varepsilon\rightarrow0$.
The left-hand side of (\ref{linear-combination-x})
consists of $\varepsilon$-divergent terms
and contains no $\varepsilon$-independent terms,
so that
we conclude that
the surface term in (\ref{level-x}) 
never contribute to the renormalized inner product
defined in (\ref{gamma-product}).
The surface term in (\ref{level-y}) has the same $\varepsilon$-behavior
as that in (\ref{level-x}),
because
the surface term 
$
  \ST_y
  \left[ \langle\phi_{n_+,-m_-},\:J\:\phi_{n'_+,-m'_- + 1}\rangle
  \right]_\varepsilon
$
is obtained by a linear combination of the functions
$x^a y^b e^{-x^2 + y^2}$,
and we have a relation
\begin{eqnarray}
 \ST_y
  \left[ x^a y^b e^{-x^2 + y^2} \right]_\varepsilon
  &=&
  \frac{2}{-\log\varepsilon}
  \:
  \left(\frac{1}{\varepsilon}\right)^{\frac{a+b+1}{2}}
  \: \times
  \left\{
   \begin{array}{ll}
     1 & (a=0,\ b=1)\\
     \frac{a+b-1}{2}\Gamma\left(\frac{a+b-1}{2}\right)
       & (a:{\rm even},\ b:{\rm odd})\\
     0 & ({\rm others})
   \end{array}
  \right..
\end{eqnarray}
Thus the surface term in (\ref{level-y}) has no contribution
to the renormalized inner product (\ref{gamma-product}).

As we have seen above,
the surface terms in
the recurrence formulae (\ref{level-x}) and (\ref{level-y})
consist of divergent terms whose $\varepsilon$-dependence is
\begin{eqnarray}
 &&
 \frac{1}{-\log\varepsilon} \left(\frac{1}{\varepsilon}\right)^n
 \qquad
 (n:\ \mbox{positive integer}).
 \label{epsilon-dependence}
\end{eqnarray}
When we multiply (\ref{epsilon-dependence}) by $\varepsilon$,
the product never contain $\varepsilon$-independent part.
Thus the third term in (\ref{level-x}) never contribute
to the renormalized inner product (\ref{gamma-product}).

Finally
the recurrence formulae (\ref{level-x}) and (\ref{level-y})
tell us that $\varepsilon$-regularized inner product
(\ref{pre-gamma-product}) has the property:
\begin{eqnarray}
  \langle n_+, -m_- \:|\: n_+', -m_-'  \rangle_\varepsilon
  &=&
  (\varepsilon\mbox{-divergent part})
  \;+\;
  \delta_{n_+,n_+'}
  \delta_{m_-,m_-'}
  \;+\;
  (\varepsilon\mbox{-zero part}),
  \label{resultant-separation}
\end{eqnarray}
and we conclude that the renormalized inner product
(\ref{gamma-product})
satisfies the orthonormal condition:
\begin{eqnarray}
  \langle n_+, -m_- \:|\: n_+', -m_-'  \rangle
  &=&
  \delta_{n_+,n_+'}
  \delta_{m_-,m_-'}.
\end{eqnarray}

\section{Holomorphic Regularization without Subtraction}

One may consider that
the separation of the divergent part in (\ref{separation})
has ambiguity
and the definition of the renormalization (\ref{gamma-product})
is also ambiguous.
As we have seen in the proof in the Section 3,
there is no ambiguity because the structure of the divergent part
is completely understood.
In this section,
we present another definition of the inner product
which obviously has no ambiguity
because the subtraction scheme is not employed.

We extend the regularization parameter $\varepsilon$
in (\ref{pre-gamma-product}) from real value to complex one.
The definition of the inner product is
\begin{eqnarray}
 \langle f \:|\: g \rangle
  &\equiv&
 \lim_{\alpha \to 0}
 \left[
 \lim_{\varphi \to \infty}
 \langle f \:|\: g \rangle_{\varepsilon = \varepsilon(\alpha, \varphi)}
 \right],
 \label{holomorphic-product}
\end{eqnarray}
where we have introduced the parameterization of $\varepsilon$ as
\begin{eqnarray}
 \varepsilon(\alpha, \varphi)
  &\equiv& \alpha \; e^{i \varphi},
\end{eqnarray}
namely, the positive parameter $\alpha$ is the absolute value,
and the real parameter $\varphi$ is the phase.
The limit of the phase $\varphi$ in (\ref{holomorphic-product})
is continuously taken according to the path
\begin{eqnarray}
 \varphi &=& 0 \to 2\pi \to 4\pi \to 6\pi \to \cdots \to \infty \times 2\pi.
 \label{phi_path}
\end{eqnarray}
The absolute value $\alpha$ should be always smaller than 1 in the limit.
The order of the limit in (\ref{holomorphic-product}) is quite important,
namely,
the limit of $\varphi$ should be taken before the limit of $\alpha$.
There is no subtraction in the definition (\ref{holomorphic-product}),
thus there never arises ambiguity in the inner product.

We will proof the orthonormality 
of the inner product (\ref{holomorphic-product}).
First we confirm that
the norm of the vacuum $|0_+,0_- \rangle$ becomes unity.
By recalling the relation (\ref{pre-vacuum-norm}),
the vacuum norm becomes
\begin{eqnarray}
  \langle 0_+,0_-|0_+,0_- \rangle
   &=&
   \lim_{\alpha \to 0}
   \left[
   \lim_{\varphi \to \infty}
   \langle 0_+,0_-|0_+,0_- \rangle_{\varepsilon(\alpha,\varphi)}
   \right]
   \nonumber\\
  &=&
   \lim_{\alpha \to 0}
   \left[
   \lim_{\varphi \to \infty}
   \frac{1}{\sqrt{\varepsilon +1}}
   \times
   \frac{
    \log\left(\sqrt{\varepsilon +1}+1\right)
   -\log\left(\sqrt{\varepsilon +1}-1\right)}{-\log\varepsilon}
   \right].
   \label{vacuum-norm-2a}
\end{eqnarray}
By taking the path (\ref{phi_path}),
the first factor $1/\sqrt{\varepsilon +1} = 1/\sqrt{\alpha e^{i\varphi}+1}$
in (\ref{vacuum-norm-2a})
has no extraordinary contribution,
because $|\alpha| < 1$ and
the path (\ref{phi_path}) does not cross the branch-cut of the square root.
In the second factor of (\ref{vacuum-norm-2a}),
while the pass never crosses the branch cut of
$\log\left(\sqrt{\varepsilon +1}+1\right)$ in the numerator,
the pass go across the branch cut of $\log\varepsilon$ in the denominator
and of $\log\left(\sqrt{\varepsilon +1}-1\right)$ in the numerator.
Therefor
we find that the vacuum norm (\ref{vacuum-norm-2a})
becomes unity as
\begin{eqnarray}
  \langle 0_+,0_-|0_+,0_- \rangle
  &=&
   \lim_{\alpha \to 0}
   \frac{1}{\sqrt{\alpha +1}}
   \left[
   \lim_{\varphi \to \infty}
   \frac{\log\left(\sqrt{\alpha +1}+1\right)           }{-(\log\alpha) - i\varphi}
   -
   \frac{\log\left(\sqrt{\alpha +1}-1\right) + i\varphi}{-(\log\alpha) - i\varphi}
   \right]
   \nonumber \\
  &=&
   1.
   \label{vacuum-norm-2}
\end{eqnarray}

According to the argument in the previous section,
the ($\varepsilon$-divergent part) in (\ref{resultant-separation})
consists of the terms in (\ref{epsilon-dependence}),
and the ($\varepsilon$-zero pert) consists of the terms
\begin{eqnarray}
 &&
 \frac{1}{-\log\varepsilon} \varepsilon^n
 \qquad
 (n:\ \mbox{positive integer}).
 \label{epsilon-zero-terms}
\end{eqnarray}
By taking the limit according to the path (\ref{phi_path}),
the factor $\varepsilon^{\pm n}$
in the ($\varepsilon$-divergent part) and
in the ($\varepsilon$-zero pert) 
has no additional contribution
because $n$ is a integer,
and the only logarithm in the denominator produces
the additional phase factor $i\varphi$.
Thus, these terms behave in the limit as
\begin{eqnarray}
 \lim_{\alpha \to 0}
  \left[
   \lim_{\varphi \to \infty}
   \frac{1}{-\log\varepsilon(\alpha,\varphi)}
   \varepsilon(\alpha,\varphi)^{\pm n}
  \right]
 &=&
 \lim_{\alpha \to 0}
  \left[
   \lim_{\varphi \to \infty}
   \frac{1}{-(\log\alpha) - i \varphi}
  \right]
   \alpha^{\pm n}
   \nonumber\\
 &=&
  0,
\end{eqnarray}
where $n$ is a positive integer.
We have found that
the ($\varepsilon$-divergent part)
and the ($\varepsilon$-zero pert)
vanish in the limit of $\varphi$.
Therefore,
the inner product defined in (\ref{holomorphic-product})
results the orthonormal relation:
\begin{eqnarray}
  \langle n_+, -m_- \:|\: n_+', -m_-'  \rangle
   &=&
   \lim_{\alpha \to 0}
   \left[
   \lim_{\varphi \to \infty}
   \langle n_+, -m_- \:|\: n_+', -m_-' \rangle_{\varepsilon(\alpha,\varphi)}
   \right]
   \nonumber\\
  &=&
  \delta_{n_+,n_+'}
  \delta_{m_-,m_-'}.
\end{eqnarray}


\section{Conclusion and future perspectives}

We have proposed the other definitions of the
positive definite inner product for the negative number sector
of the double harmonic oscillator
than the non-local method in the previous paper \cite{Habara:2006eg}.
The new definitions provide positive definite inner product,
and there arises no difference of results among the definitions.

The separation of $(\varepsilon\mbox{-independent part})$
from $(\varepsilon\mbox{-divergent part})$ in (\ref{separation})
has been succeeded,
because we have known the form of 
$(\varepsilon\mbox{-divergent part})$
by the concrete calculation.
One may consider that
the separation in (\ref{separation}) may seem to be ambiguous,
however,
we have found that
the anxiety of the ambiguity is completely solved by
the holomorphic definition (\ref{holomorphic-product}).
This means that there arises no ambiguity in the perturbation theories,
because the perturbation theories never deform the harmonic oscillators
in the mode expansion of the fields.
When we consider the non-perturbative phenomena, i.g.,
quantum anomalies in external fields,
the ambiguity may arises and results the non-trivial effects.
This is the further subject to investigate.

\section*{Acknowledgement}

We acknowledge R.~Jackiw for his encouraging communication who shares
with us the view point that the Dirac sea method provides deep physical
understanding and intuition to some novel phenomena.
We also thank T.~Iwaki, Z.~Tsuboi and T.~Tsuchida
for useful discussions.
This work is supported by Grants-in-Aid for Scientific Research on
Priority Areas, Number of Areas 763, ``Dynamics of Strings and Fields'',
from the Ministry of Education of Culture, Sports, Science and
Technology, Japan.


\appendix
\section{APPENDIX}

We present a detailed derivation of (\ref{dif-x}) in the following:
\begin{eqnarray}
 &&
  \left< f \:|\: {\textstyle\frac{d}{dx}g} \right>_\varepsilon
  \nonumber\\
  &=&
  \int_{\gamma}dxdy \: f \cdot \frac{d}{dx}g \cdot e^{-\varepsilon x^2}
  \nonumber\\
  &=&
  \int_{-\infty}^{+\infty} dr
  \int_{-\infty}^{+\infty} d\theta \:
  |r| f \cdot
  \left\{\cosh \theta \frac{d}{dr}
   -\frac{\sinh \theta}{r}\frac{d}{d\theta}
  \right\} g \cdot
  e^{-\epsilon r^2 \cosh^2 \theta}
  \nonumber\\
 &=&
  \left[
   \int_{-\infty}^{+\infty} d\theta \: |r| \cosh\theta \cdot fg
   \cdot e^{-\epsilon r^2 \cosh^2 \theta}
  \right]_{r=-\infty}^{r=+\infty}
  \nonumber\\
 && \qquad
 -
  \left[
   \int_{-\infty}^{+\infty} dr \: \frac{|r|}{r} \sinh\theta \cdot fg
   \cdot e^{-\epsilon r^2 \cosh^2 \theta}
  \right]_{\theta =-\infty}^{\theta =+\infty}
  \nonumber\\
 && \qquad
  -\int_{-\infty}^{+\infty} dr
   \int_{-\infty}^{+\infty} d\theta \:
   \left[
     \left\{\frac{d}{dr} \left(|r| f\right) \right\} \cosh\theta \cdot g
    -\frac{|r|}{r}\left\{\frac{d}{d\theta} \left(f\sinh\theta\right)\right\}g
   \right] \cdot e^{-\epsilon r^2 \cosh^2 \theta}
   \nonumber \\
 && \qquad
  -\int_{-\infty}^{+\infty} dr
   \int_{-\infty}^{+\infty} d\theta \:
   |r| f g 
   \left\{
     \cosh\theta \cdot  \frac{d}{dr}
    -\frac{\sinh\theta}{r} \frac{d}{d\theta} 
   \right\} e^{-\epsilon r^2 \cosh^2 \theta}
   \nonumber \\
 &=&
  [\mbox{surface terms}] \nonumber \\
 && \qquad
  -\int_{-\infty}^{+\infty}dr
   \int_{-\infty}^{+\infty}d\theta \:
   |r|
   \left[\left\{\cosh \theta \frac{d}{dr}f\right\}
    -\left\{\frac{\sinh \theta}{r}\frac{d}{d\theta}f\right\}
   \right] \cdot g
   \cdot e^{-\epsilon r^2 \cosh^2 \theta} 
  \nonumber \\
 && \qquad
  -\int_{-\infty}^{+\infty} dr
   \int_{-\infty}^{+\infty} d\theta \:
   |r| f g 
   \left(-2 \varepsilon \right)
   r \cosh \theta \:
   e^{-\epsilon r^2 \cosh^2 \theta}
   \nonumber \\
 &=&
  [\mbox{surface terms}]
  \;-\;
  \int_{\gamma} dxdy \: \frac{d}{dx}
  f \cdot g
  \cdot e^{-\epsilon x^2 }
  \;+\; 2 \varepsilon
  \int_{\gamma} dxdy \: f \cdot g \: x
  \cdot e^{-\epsilon x^2 }
  \nonumber\\
 &=&
  [\mbox{surface terms}]
  \;-\;
  \left< {\textstyle\frac{d}{dx}f} \:|\: {g} \right>_\varepsilon
  \;+\; 2 \varepsilon
  \left< {f} \:| x |\: {g} \right>_\varepsilon.
  \label{h.2.12}
\end{eqnarray}
%


\end{document}